\documentclass{moriond}

\def\be{\begin{equation}}
\def\ee{\end{equation}}
\def\bea{\begin{eqnarray}}
\def\eea{\end{eqnarray}}

\newcommand{\Photo}{}

\begin{document}
\vspace*{4cm}
\title{On-going improvements in the Virgo strain h(t) reconstruction, online noise subtraction and early-warning pipe in preparation for the O4 run}

\author{Monica Seglar-Arroyo on behalf of the Virgo Collaboration}

\address{Institut de Física d’Altes Energies (IFAE), BIST, Barcelona, Spain}

\maketitle\abstracts{
In this contribution, we outline the improvements on the strain h(t) reconstruction in preparation for the observing run O4 of AdvancedVirgo+. These improvements have the main goal to provide a h(t) with high precision and reduce its uncertainties on this quantity to a few percent. First, we describe how the reconstruction of the strain signal h(t) is performed in Virgo and its link to the interferometer and its calibration. We highlight how we plan to monitor the optical response of the interferometer and mirrors. We will describe how we plan to correct the bias of the reconstructed h(t) strain. We present the new online linear noise subtraction method, developed to successfully tackle correlated noise witness channels that are also present in h(t). We provide the status of the low-latency h(t) strain reconstruction, which has the main goal to reduce the latency in pre-merger early warning alerts.}

\section{The reconstruction of the gravitational wave strain h(t) in O4 }
The AdvancedVirgo+ detector is a double-recycled Michelson interferometer sensible to gravitational waves in the 10$\,$Hz-10$\,$kHz band, which is part of the LIGO-Virgo-KAGRA network. Since the discovery of gravitational waves in 2015, 90 compact binary coalescences have been detected~\cite{1}. Various improvements in the reconstruction of the strain signal h(t) have been foreseen for the next observing run, O4. The length variations of the arms of the interferometer caused by the passage of a gravitational wave are observed as power variations in the output photodiode signal. However, the ITF is controlled via control loops to be kept on a working point to optimise the sensitivity, so the control loops dominate the output signals in the 10$\,$Hz -100$\,$Hz band, i.e. part of the GW signal is in the control signals. The reconstruction of h(t)~\cite{2} consists of the combination of the output signal, control signals and witness channels, with the knowledge on the response of the mirror/marionettes actuators and the knowledge of the mirror and interferometer optical response, to go back to the gravitational wave strain h(t). The derived frequency-independent uncertainties during O3, in the 20–2000 Hz band, are of 5\% in amplitude, 35 mrad in phase and 10 $\mu$s in timing (with larger uncertainties around 50 Hz).

There are two h(t)-reconstruction streams planned for O4 in Virgo. The main stream is the low-latency h(t) stream, provided in $\sim$10 seconds. It includes the adjustment of the complex optical response models of the mirrors and the interferometer, linear noise subtraction, correction of the bias observed in the $\sim$20$\,$Hz-500$\,$Hz band, regular update of uncertainties on online h(t) (goal being modulus $<$$\,$3$\,$\% and phase $<$20$\,$mrad), and the distribution of h(t) in files with updated uncertainties for offline analysis. Then, an early-warning h(t) stream with the goal to issue alerts before the merger of the binary neutron star, for which the latencies of the reconstruction have been minimised to $\sim$5 seconds (from the shortening by half of the fast Fourier transform size, since it is the main driver of the latency in the reconstruction process). The uncertainty on the gravitational wave strain h(t) is computed as the quadratic sum of various contributions: the uncertainties on the actuator models, obtained from internal calibrators (electromagnetic actuators) and external calibrator (photon calibrator~\cite{3} and newtonian calibrator~\cite{4}), those of the sensing and timing of the output photodiode and the temporal stability. The monitoring of the h(t) reconstruction uncertainties is obtained via injections with high SNR at large number of frequencies (from $\sim$10$\,$Hz to 1500$\,$Hz), once per week during out of observing mode periods, and via continuous monitoring, which consists on 12 permanent low-SNR sinusoidal lines in the most sensitive band of the ITF $\sim$35$\,$Hz-400$\,$Hz. Note that, while in O3 these were given as frequency-independent uncertainties, computed for 6-month data chunks, the goal for O4 is to provide frequency-dependent uncertainties, computed on a weekly basis.

\subsection{Optical response adjustment in the low-latency h(t) reconstruction}
The optical response of the mirrors and the interferometer, in W/m, characterises how the length variations due to movement of mirrors produce variations in the output signal. Calibration lines are continuously injected to characterise and follow the optical response of the interferometer and mirrors in real time. The parametric functions describing the interferometer optical response of AdvanceVirgo+ can be given with two \textit{features} due to the newly installed Signal Recycling mirror: an optical spring ($\sim$20$\,$Hz-30$\,$Hz) and a double cavity pole ($\sim$350$\,$Hz-400$\,$Hz). The number of injected lines depend on the number of parameters of the model. These are injected around the frequencies where the features are expected.
      
\subsection{Linear noise subtraction of correlated noises in the low-latency h(t) reconstruction}
Noise contribution from many sources, e.g. laser frequency noise or scattered light noise, can couple to the h(t) signal. Witness channels are selected to monitor these contributions to linearly subtract them from h(t) via transfer functions. However, these noises can be observed by several NWCh. To tackle this issue, a noise subtraction method of uncorrelated transfer functions  between h(t) and the noise witness channel~\cite{5}  has been adapted to Virgo. First results with simulated noise injections in O3 data, with up to 5 NWCh observing 5 broadband coherent noises with h(t) and coherent among them, show significant improvements starting from 20\% of coherence between NWCh. For large coherences between the NWCh and h(t), this noise subtraction improves by $\sim$25\% the BNS range achieved. The largest improvement comes from the flexibility of the frequency range selection of the NWCh and from the protection it provides against divergencies in h(t) observed in O3 when more than 2 NWCh observed the same noise.

%
%

\section*{Acknowledgments}

This work is partially supported by the Spanish MCIN/AEI/10.13039/501100011033 under the Grants No. SEV-2016-0588, No. PGC2018-101858-B-I00, and No. PID2020-113701GB-I00, some of which include ERDF funds from the European Union, and by the MICIIN with funding from the European Union NextGenerationEU (PRTR-C17.I1) and by the Generalitat de Catalunya. MSA acknowledges the support of the Grant FJC2020-044895-I. IFAE is partially funded by the CERCA program of the Generalitat de Catalunya. LVK acknowledgements may be found in https://dcc.ligo.org/P2100218


\section*{References}

\begin{thebibliography}{99}
\bibitem{1} Abbott, R., et al., arXiv:2111.03606 (2021).
\bibitem{2} Acernese, F., et al.,Classical and Quantum Gravity 39.4 (2022): 045006.
\bibitem{3} Estevez, D., et al., Classical and Quantum Gravity 38.7 (2021): 075007.
\bibitem{4} Estevez, D., et al., Classical and Quantum Gravity 38.7 (2021): 075012.
\bibitem{5} Davis, Derek, et al., Classical and Quantum Gravity 36.5 (2019): 055011.

\end{thebibliography}
\end{document}